\documentclass{aa}  
\usepackage{graphicx}
\usepackage{txfonts}

\titlerunning{Discovery of an LSB galaxy in the ZoA with VVVX}
\authorrunning{J. L. Nilo-Castell\'on et al.}

\begin{document} 

\title{Discovery of a low surface brightness galaxy in the Zone of Avoidance using VVVX Deepstack imaging}

\author{J. L. Nilo-Castell\'on\inst{1} \and  L. Baravalle\inst{2,3}  \and D. Minniti\inst{5,6} \and  M. V. Alonso\inst{2,3} \and  G. Galaz\inst{7} \and  C. Villalon\inst{2,3} \and  R. Hinojosa\inst{9} \and  E. O Schmidt\inst{2,3} \and C. Valotto\inst{2,3} \and M. Soto\inst{4} \and E. G. Rivera\inst{1} \and  V. Valdivia-Rojas\inst{1} \and D. Alvarez-Rocha\inst{1} \and P. Marchant Cortés\inst{1}  \and N. J. G. Cross\inst{8}}  

\institute{  
Departamento Astronom\'ia, Facultad de Ciencias, Universidad de La Serena. Av. Raul Bitran 1305, La Serena, Chile.
\and
Instituto de Astronom\'{\i}a Te\'orica y Experimental, (IATE-CONICET), Laprida 854, X5000BGR, C\'ordoba, Argentina.
\and
Observatorio Astron\'omico de C\'ordoba, Universidad Nacional de C\'ordoba, Laprida 854, X5000BGR, C\'ordoba, Argentina.
\and
Instituto de Investigación en Astronomía y Ciencias Planetarias, Universidad de Atacama, Av. Copayapu 485, Copiapó, Chile.
\and
 Instituto de Astrof\'isica, Facultad de Ciencias Exactas, Universidad Andr\'es Bello, Av. Fernandez Concha 700, Las Condes, Santiago, Chile.
 \and
Vatican Observatory, V00120 Vatican City State, Italy.
\and
Instituto de Astrofísica, Pontificia Universidad Católica de Chile, Vicuña Mackenna 4860, 7820436 Macul, Santiago, Chile
\and
Wide-Field Astronomy Unit, Institute for Astronomy, University of Edinburgh, Royal Observatory, Blackford Hill,
Edinburgh EH9 3HJ, UK
\and
Cerro Tololo Inter-American Observatory, CTIO/AURA Inc., La Serena, Chile
}   \date{Received April 13, 2026; accepted xxx xx, xxxx}

 
\abstract
{We report the discovery of VVVX\,J080705.96$-$273823.6, the first low-surface-brightness galaxy candidate identified in the Zone of Avoidance by the VISTA Variables in the V\'ia L\'actea eXtended Survey (VVVX). The source was detected serendipitously during a visual inspection of the new Deepstack images, a set of reconstructed mosaics generated through an updated reprocessing and stacking procedure that significantly improves both the depth and spatial uniformity of the original survey images. 

First-order morpho-photometric parameters were obtained using \texttt{SExtractor} in combination with PSF models derived with \texttt{PSFEx}, applied to star-subtracted images in order to minimise foreground contamination. Final structural properties of the galaxy were derived from PSF-convolved modelling of azimuthally averaged surface-brightness profiles in the $J$, $H$, and $K_{\rm s}$ bands using exponential, S\'ersic, and composite models.

The galaxy candidate is clearly detected in the three available near-infrared bands. The azimuthally averaged profiles are consistently described by a disk-dominated (exponential) model in all three bands, with exponential scale lengths in the range $h \sim 32$--36\arcsec\ and extinction-corrected central surface brightness values spanning $\mu_0 \simeq 19.82$--22.35 mag\,arcsec$^{-2}$. The $J$ band gives the best-constrained profile; $H$ and $K_{\rm s}$ show the same behaviour with larger uncertainties.

The projected proximity to the galaxies ESO\,494$-$25 and ESO\,494$-$26 raises the possibility of a physical association. Both ESO galaxies have radial velocities consistent with distances of $\sim 11$--$12$ Mpc. If VVVX\,J080705.96$-$273823.6 were located at the same distance, its characteristic exponential scale length of $h \simeq 34''$ would correspond to $\sim 1.8$--$2.0$ kpc, consistent with the range of scale lengths observed in nearby LSB disk galaxies. However, in the absence of a spectroscopic redshift, this association remains tentative.

 This discovery demonstrates that the new VVVX \textit{Deepstack} products can reveal faint diffuse galaxies at low Galactic latitudes, opening a path toward a more complete census of the obscured nearby Universe.}

\maketitle
%

\section{Introduction}
\begin{figure*}
   \centering
   \includegraphics[width=0.33\textwidth]{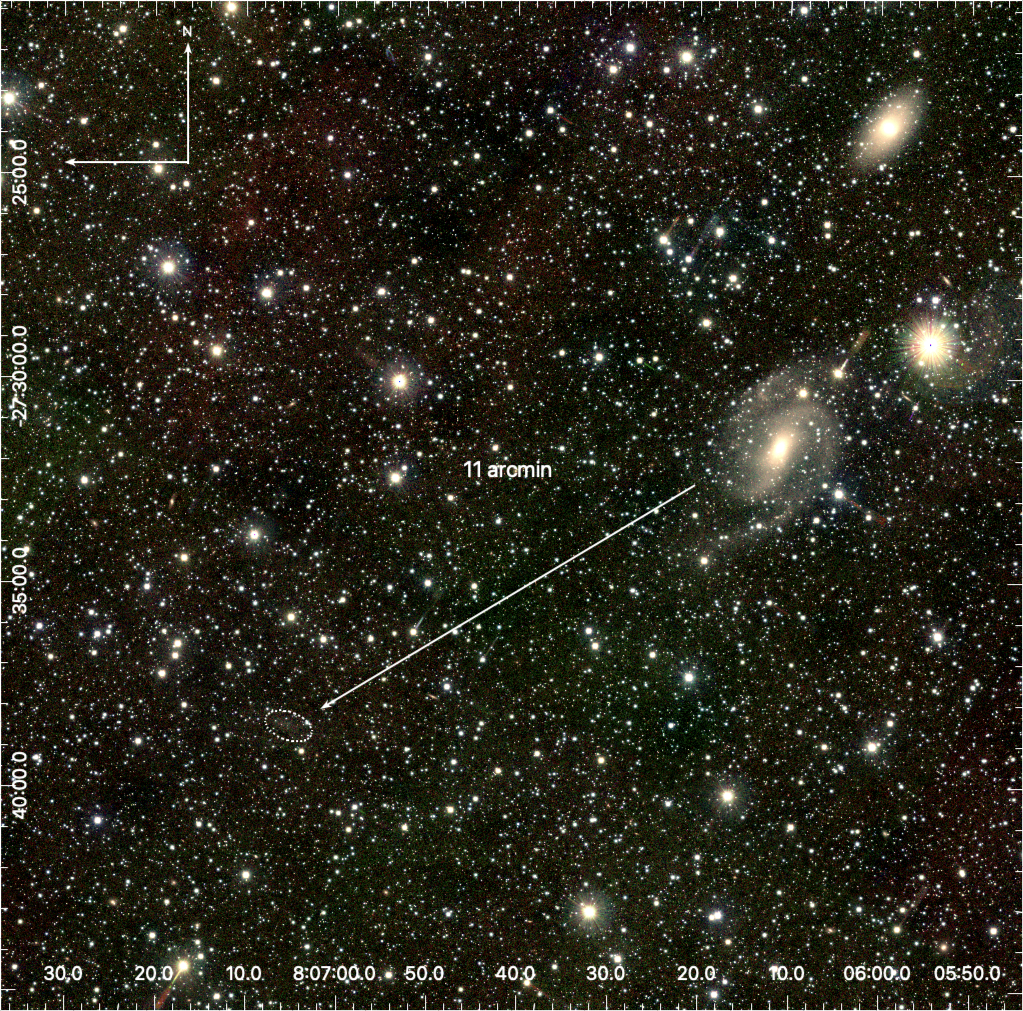}
   \includegraphics[width=0.33\textwidth]{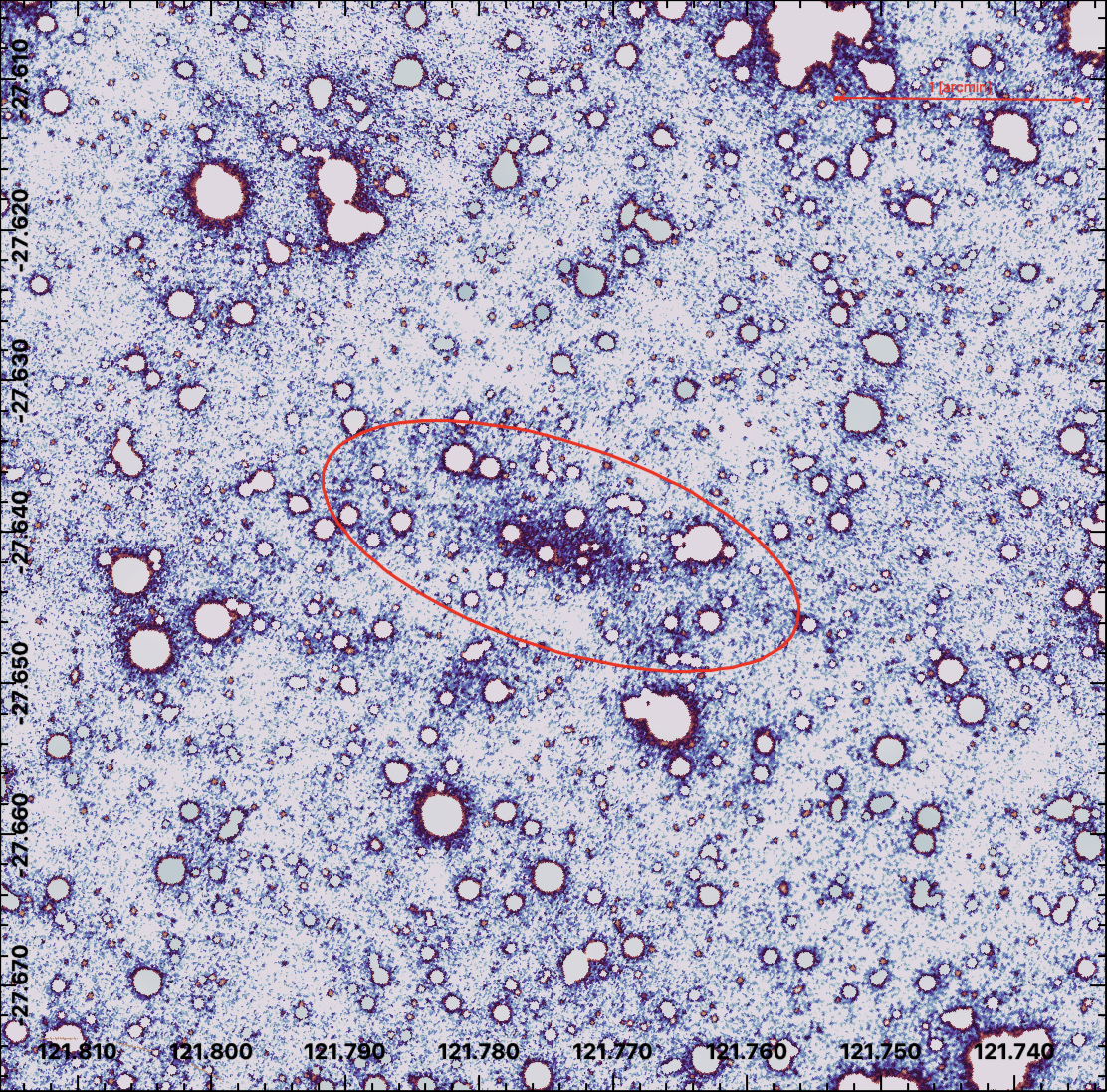}
   \includegraphics[width=0.33\textwidth]{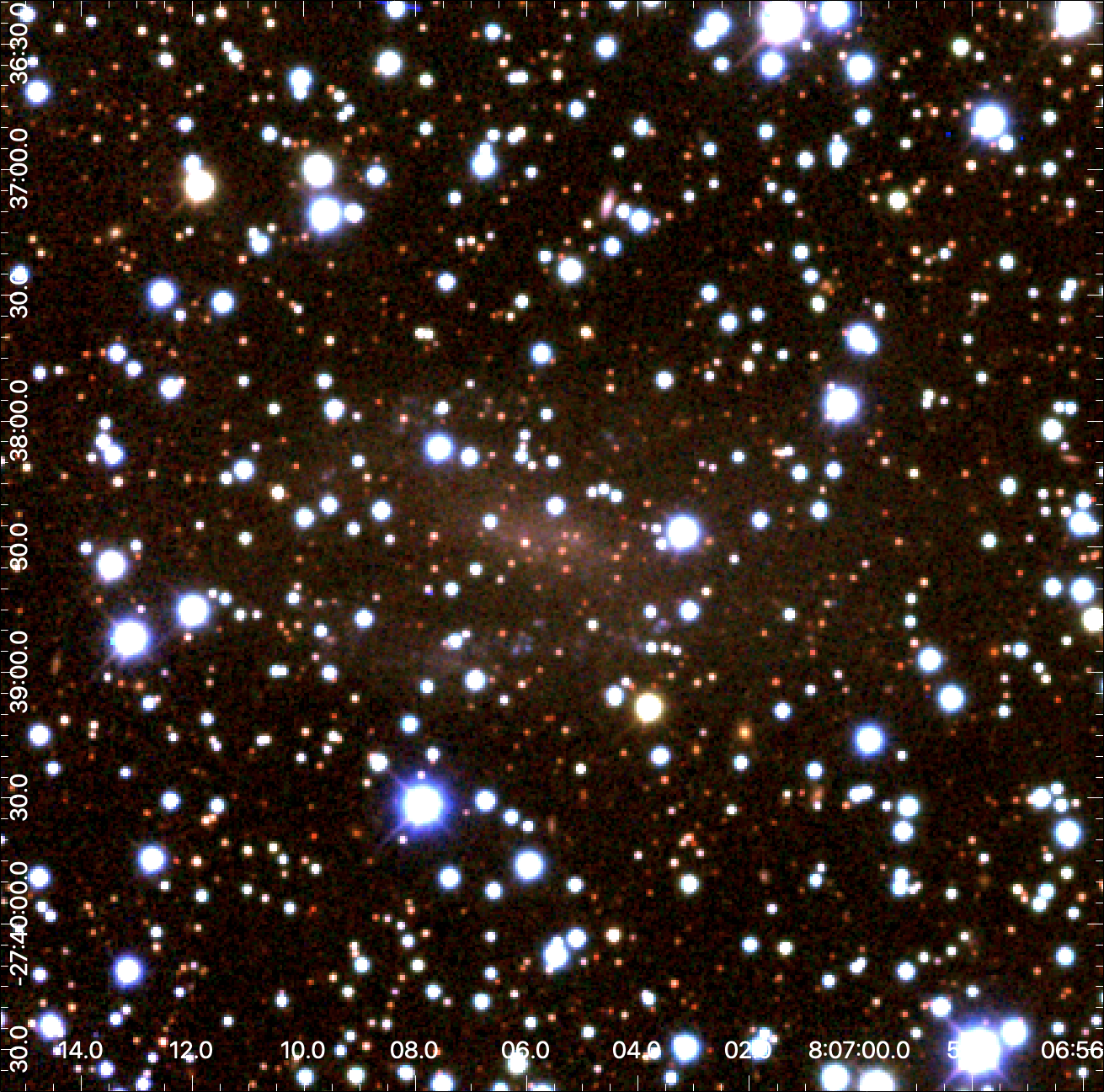}
\caption{\textbf{Left:} RGB colour-composite image constructed from a section of the VVVX \textit{Deepstack} $J$, $H$, and $K_{\rm s}$ images tile e0778. The LSB is marked by the dashed ellipse in the lower-left corner 11 arcmin in the southwest direction of the galaxy ESO\,494$-$256. \textbf{Center:} Stellar-subtracted $J$-band image of VVVX\,J080705.96$-$273823.6. The red ellipse marks the \texttt{MAG\_AUTO} aperture returned by \texttt{SExtractor}. \textbf{Right:} RGB colour-composite image constructed from the DECaPS $g'$, $r'$, and $i'$ images. Center and right panels cover a side length of $4.5'$. 
}
\label{fig1}
\end{figure*}

The Zone of Avoidance \citep[ZoA;][]{KraanKorteweg2000} remains the main observational discontinuity in attempts to reconstruct the local cosmic density field. Severe obscuration from the Milky Way, produced by dust extinction, stellar crowding, and diffuse emission \citep{Schlegel1998, Schlafly2011, Soto2019}, interrupts the continuity of large-scale structure mapping across the Galactic plane. Although extensive multi-wavelength surveys over the past decades have identified thousands of previously hidden galaxies behind the Milky Way \citep[e.g.,][]{Jarrett2000, Donley2005, Henning2010, StaveleySmith2016, KraanKorteweg2018, Baravalle2018, Baravalle2021, Daza2023, Galdeano2023, Alonso2025, Baravalle2026}, these efforts have mainly probed the bright end of the luminosity function. Consequently, the faint, diffuse, and low-surface-brightness galaxy population remains poorly constrained. While such systems are not expected to dominate the mass budget individually, their systematic omission can bias the tracer population used to map large-scale structure. In particular, missing faint and diffuse galaxies may lead to an incomplete sampling of structures crossing the Galactic plane, with consequences for dynamical reconstructions, peculiar-velocity fields \citep{Maller2003}, and cosmological inferences based on large-scale structure \citep{Clocchiatti2024}. A more complete census of faint and low-surface-brightness galaxies in the ZoA is therefore required for a physically consistent reconstruction of the local density field.

In this Letter, we report the discovery of VVVX\,J080705.96$-$273823.6, the first low-surface-brightness galaxy identified in the Zone of Avoidance using data from the VISTA Variables in the V\'ia L\'actea eXtended survey (VVVX). The paper is organized as follows. Sect.~\ref{sec:data} describes the data and photometric procedures, Sect.~\ref{sec:structural_analysis} presents the structural characterization, and Sect.~4 summarizes the results and discusses the possible association with nearby galaxies.

\section{The Data} \label{sec:data}

This work utilizes a new set of deep near-infrared mosaics, hereafter referred to as \textit{Deepstack}, reconstructed from the VISTA Variables in the V\'ia L\'actea eXtended survey \citep[VVVX;][]{VVVX}. Updated stacking procedures and revised calibrations were applied to the original VVVX images, producing mosaics with increased depth and improved spatial uniformity\footnote{The \textit{Deepstack} images are available through the VISTA Science Archive (VSA; \url{https://vsa.roe.ac.uk}) for the VVVX team. }. 

Our analysis focuses on the \textit{Deepstack} images of tile e0778 in the $J$, $H$, and $K_{\rm s}$ bands. The stacks combine 36, 24, and 264 individual frames, yielding total integration times of 1080\,s ($J$), 288\,s ($H$), and 1056\,s ($K_{\rm s}$). The corresponding 5$\sigma$ limiting magnitudes for point sources are $J = 20.61$, $H = 19.35$, and $K_{\rm s} = 18.80$ mag (Vega). All mosaics preserve a pixel scale of $0\farcs34$. We estimated the limiting surface brightness from the sigma-clipped background dispersion in source-free regions of the star-subtracted images, following \citet{Roman2020}. Over $10''\times10''$ areas we obtain $3\sigma$ limits of $\mu_{\rm lim}\simeq25.6$, $24.3$, and $23.6$ mag\,arcsec$^{-2}$ in $J$, $H$, and $K_{\rm s}$, the $J$ band being the deepest.

We measured first-order structural parameters with \texttt{SExtractor} in a $7' \times 7'$ region centred on VVVX\,J080705.96$-$273823.6. To minimise contamination from the dense stellar foreground while preserving the galaxy's diffuse emission, we adopted a two-pass procedure. In the first pass, \texttt{SExtractor} was run with a configuration optimised for compact-source detection, using a Mexican-hat filter, \texttt{DETECT\_THRESH}=2, and a large background mesh of \texttt{BACK\_SIZE}=192 pixels. The compact sources detected in this pass were used to define a foreground-source mask and to generate a foreground-subtracted image. In the second pass, \texttt{SExtractor} was applied to the foreground-subtracted image with a configuration optimised for faint extended emission, using a Gaussian filter, \texttt{DETECT\_THRESH}=1, and an area-based selection to isolate the galaxy.

\begin{table}[h]
\centering
\small
\begin{tabular}{lccccc}
\hline
Filter & N stars & $\chi^{2}$ & FWHM (px) & ellipticity & asym. \\
\hline
$J$          & 210 & 0.39 & 2.72 & 0.01 & 0.03 \\
$H$          & 276 & 0.42 & 2.98 & 0.01 & 0.05 \\
$K_{\rm s}$ & 281 & 0.36 & 2.51 & 0.02 & 0.03 \\
\hline
\end{tabular}
\caption{Main properties of the \textsc{PSFEx} PSF models: number of stars, reduced $\chi^{2}$, FWHM (px), ellipticity ($1-b/a$), and asymmetry.}
\label{tab:psf}
\end{table}

To assess the impact of remaining foreground-subtraction residuals, we measured the residual sky level in source-free elliptical annuli outside the main body of the galaxy. The residual background is consistent with zero, with a region-to-region systematic uncertainty of $\sim 0.2\%$ of the central intensity. Propagating this systematic through the exponential profile changes $\mu_{0}$ by $\leq 0.003~{\rm mag~arcsec^{-2}}$ and $h$ by $\leq 1\%$, and affects the outer surface-brightness profile by $\leq 0.02~{\rm mag~arcsec^{-2}}$ out to $r \sim 80''$. At radii larger than $r \sim 95''$, the profile becomes increasingly affected by subtraction residuals and was therefore excluded from the fit. The adopted fitting range is $6'' < r < 95''$. We therefore conclude that the derived structural parameters are not significantly affected by foreground-subtraction residuals.

The point-spread function (PSF) in each band was modelled with PSFEx v3.21.1 (Bertin 2013), using SExtractor detections of point-like sources selected within the stellar locus and with signal-to-noise ratios greater than 10. We adopted empirical pixel-based PSF models on a 25$\times$25 grid.  The resulting PSFs yield FWHM values of 2.72, 2.98, and 2.51 pixels in the $J$, $H$, and $K_{\rm s}$ bands, corresponding to 0.92, 1.01, and 0.85 arcsec, respectively (Table~\ref{tab:psf}).

Following \citet{nc25}, we verified the photometric calibration through comparison with 2MASS photometry retrieved using \texttt{TOPCAT} \citep{TOPCAT}. The 2MASS magnitudes were converted to the VISTA system following \citet{Soto2019}. Cross-matching with Gaia DR3 \citep{GaiaDR3} within $0\farcs5$ ensured the selection of reliable point sources (\texttt{PSS} $>0.99$). The resulting median offsets are $\Delta J = 0.02 \pm 0.10$, $\Delta H = 0.05 \pm 0.13$, and $\Delta K_{\rm s} = 0.07 \pm 0.035$ mag, consistent with the adopted zero-points of 26.2, 25.7, and 24.4 in $J$, $H$, and $K_{\rm s}$, respectivelly. All magnitudes were corrected for Galactic extinction following \citet{Schlafly2011}, adopting $A_J = 0.38$, $A_H = 0.24$, and $A_{K_{\rm s}} = 0.16$ mag. Given the $\sim 6'$ resolution of the extinction maps, a single value was adopted across the source; residual spatial variations at this low Galactic latitude cannot be excluded.

\begin{figure*}
\centering
\includegraphics[width=0.33\textwidth]{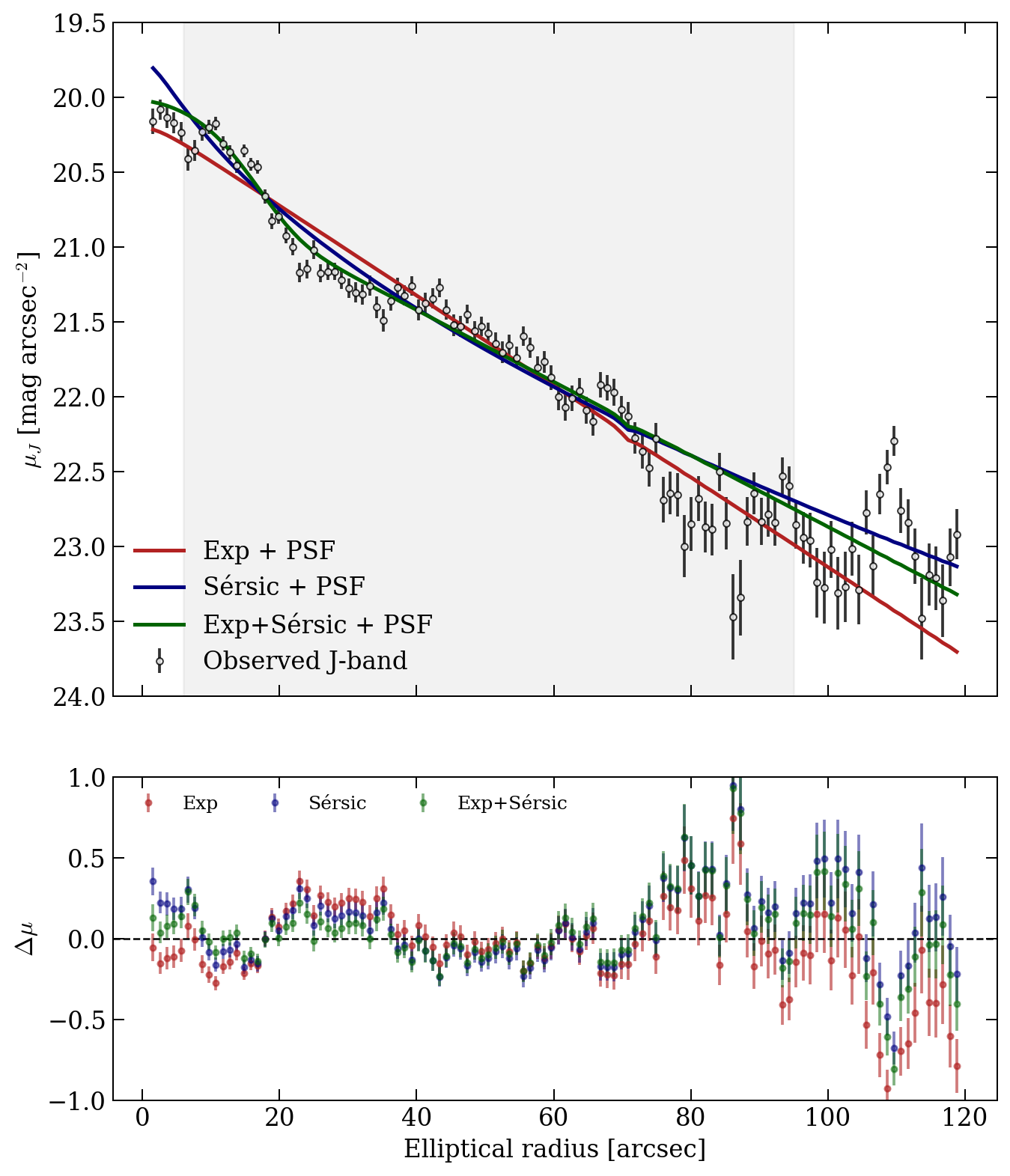}
\includegraphics[width=0.33\textwidth]{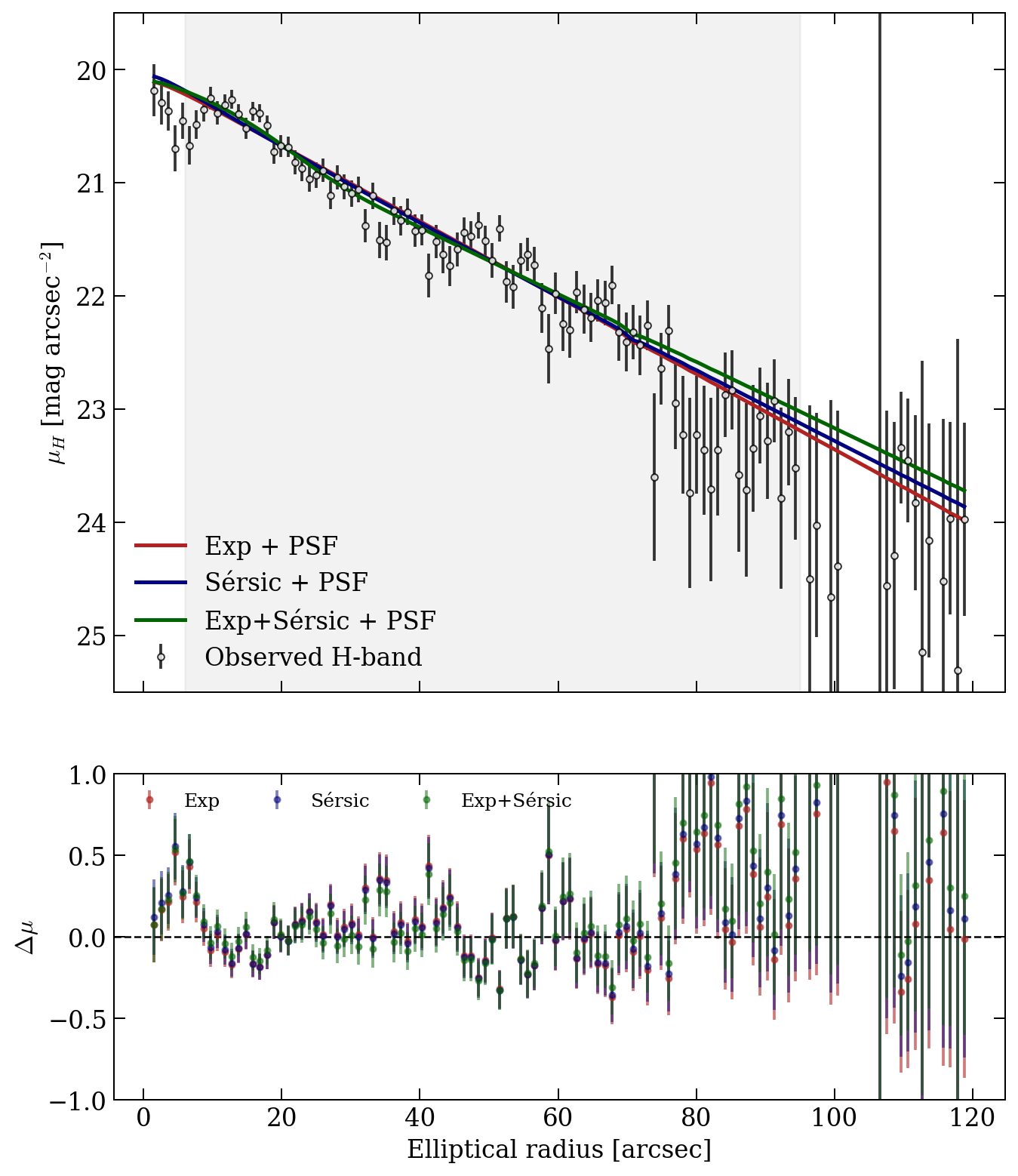}
\includegraphics[width=0.33\textwidth]{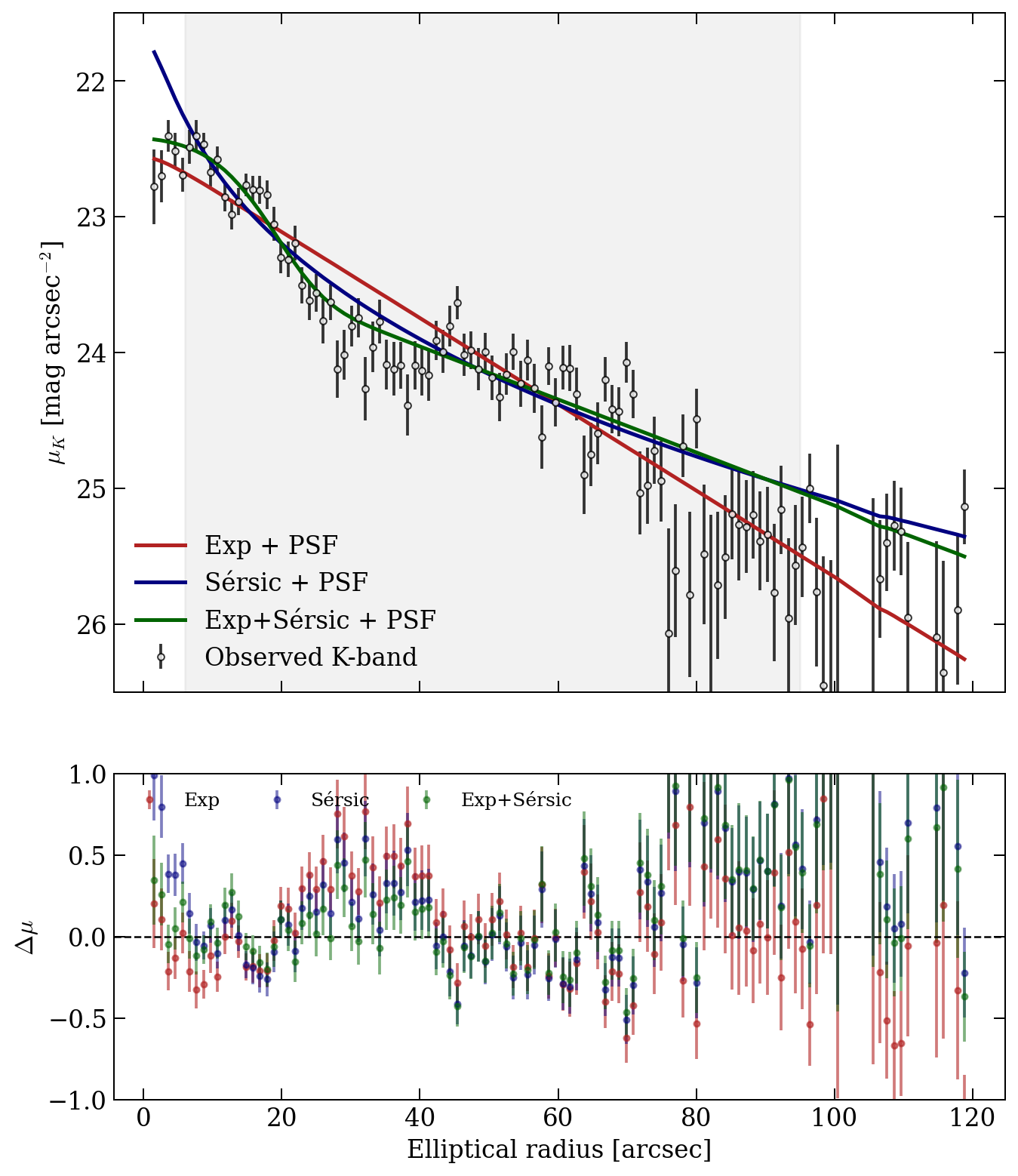}
\caption{Azimuthally averaged surface-brightness profiles of VVVX\,J080705.96$-$273823.6 in the $J$, $H$ and $K_{\rm s}$ bands. Top panels: observed profiles (black circles) and PSF-convolved exponential (red), S\'ersic (blue), and exponential+S\'ersic (green) models; shaded regions indicate the fitted radial ranges. Bottom panels: residuals relative to each model. }
\label{fig:profiles}
\end{figure*}

\section{Structural characterization of VVVX\,J080705.96$-$273823.6}
\label{sec:structural_analysis}

We detect VVVX\,J080705.96$-$273823.6 at $\alpha = 08^{\rm h}07^{\rm m}05\fs96$ and $\delta = -27^\circ38'23\farcs6$ (J2000). In the $J$-band image, the galaxy is most clearly detected as a diffuse, elongated, and spatially extended system, with its major axis oriented approximately northeast-southwest. The LSB nature of the source is also supported by optical images from the DECaPS2 survey \citep{decaps} (Fig.~\ref{fig1}). To characterize its structure, we followed a two-step approach: we first used automated source modelling to obtain initial constraints on the size, geometry, and light distribution of the galaxy, and then refined this characterization through direct modelling of the azimuthally averaged surface-brightness profiles.

First-order structural parameters were obtained with \texttt{SExtractor} using \texttt{PSFEx} PSF models, following the procedure of Sect.~\ref{sec:data}. The resulting morpho-photometric parameters are summarised in Table~\ref{tab:sextractor}, and shows a broadly consistent projected geometry in the $J$ and $H$ bands, with $b/a \sim 0.37$--$0.41$ ($\epsilon \sim 0.59$--$0.63$) and PA $\sim 68^\circ$--$70^\circ$. In $K_{\rm s}$, the source appears rounder, with $b/a \sim 0.61$ ($\epsilon \sim 0.39$), while maintaining a comparable position angle of PA $\sim 75^\circ$.

The \texttt{SExtractor} model-fitting results indicates a strongly disk-dominated structure in the $J$ and $K_{\rm s}$ bands, while the $H$-band fit appears less reliable. In the $J$ band, the fitted disk magnitude is very close to the total model magnitude, implying a negligible central spheroidal component at the depth of the data, and the solution is almost entirely disk dominated ($f_{\rm disk}=0.996$). The $K_{\rm s}$ band gives a similar result ($f_{\rm disk}=0.968$). By contrast, the $H$-band solution yields a much lower disk-to-total flux ratio ($f_{\rm disk}=0.182$), which is inconsistent with the other bands and likely reflects the lower robustness of the fit rather than a genuine structural difference. The fitted exponential disk scale lengths span $h\sim14.7''$--$24.4''$. Overall, these results provide a robust first-order description of a flattened, disk-dominated LSB system, with the $J$ band giving the most stable structural constraints for the subsequent surface-brightness analysis.

\begin{table}
\caption{\texttt{SExtractor}-based morpho-photometric parameters from the \textit{Deepstack} images. Magnitudes corrected for Galactic extinction \citep{Schlafly2011}. Scale lengths use the VIRCAM pixel scale $0\farcs34$\,pix$^{-1}$.}
\label{tab:sextractor}
\centering
\small
\begin{tabular}{lcccccc}
\hline\hline
Band & MAG$_{\rm AUTO}$ & MAG$_{\rm disk}$ & $f_{\rm disk}$ & $h$ & $b/a$ & PA \\
 & (mag) & (mag) &  & ($''$) &  & ($^\circ$) \\
\hline
$J$  & 14.719 & 14.724 & 0.996 & 16.96 & 0.366 & 69.61 \\
$H$  & 13.100 & 14.059 & 0.182 & 24.35 & 0.406 & 68.12 \\
$K_{\rm s}$ & 13.198 & 13.162 & 0.968 & 14.73 & 0.608 & 75.02 \\
\hline
\end{tabular}
\end{table}

\begin{table}
\caption{Pure-exponential PSF-convolved fits to the azimuthally averaged surface-brightness profiles. Surface brightnesses (mag\,arcsec$^{-2}$) corrected for Galactic extinction \citep{Schlafly2011}; $h$ in arcsec. S\'ersic and composite fits in the Appendix.}
\label{tab:all_models}
\centering
\small
\begin{tabular}{lccc}
\hline\hline
Band & $\mu_0$ & $h$ & $\mu_{0,\rm face-on}$ \\
\hline
$J$  & 19.819 $\pm$ 0.015 & 36.053 $\pm$ 0.433 & 20.912  \\
$H$  & 19.813 $\pm$ 0.031 & 32.443 $\pm$ 0.811 & 20.905 \\
$K_{\rm s}$ & 22.348 $\pm$ 0.032 & 34.178 $\pm$ 0.846 & 23.440  \\
\hline
\end{tabular}
\end{table}

To obtain quantitative structural parameters, we complemented this automated analysis by modelling the azimuthally averaged surface-brightness profiles directly. We derived the radial light distribution using elliptical apertures on the star-subtracted image. To ensure a consistent comparison between filters, we fixed the profile geometry to the $J$-band solution (centroid, axis ratio, and position angle). The mean intensity was measured in concentric annuli of fixed width and converted to surface brightness using the calibration of Sect.~\ref{sec:data}. To assess the stability of the profile against the background treatment, we measured the residual sky in source-free elliptical annuli well beyond the galaxy ($r \gtrsim 120''$): it is consistent with zero, with a region-to-region systematic of $\simeq 0.08$ counts\,pix$^{-1}$ ($\sim0.2\%$ of the central intensity). Propagating this systematic through the exponential profile changes $\mu_0$ by $\lesssim 0.003$ mag\,arcsec$^{-2}$ and $h$ by $\lesssim 1\%$ ($\Delta h \lesssim 0.34''$), and the outer profile by $\lesssim 0.02$ mag\,arcsec$^{-2}$ out to $r \sim 80''$, the radius out to which the profile remains essentially insensitive to the adopted background. Between $r \sim 80''$ and $r \sim 95''$ the residual-sky scatter grows but the profile is still reliably measured, whereas beyond $r \sim 95''$ it becomes increasingly affected by subtraction residuals from neighbouring sources.

We fitted the profiles with three functional forms: a pure exponential disk, a single S\'ersic profile, and a composite exponential+S\'ersic model. The fits were performed over the radial interval $6'' < r < 95''$ using weighted least squares, with uncertainties estimated from the dispersion of pixel values within each annulus. Figure~\ref{fig:profiles} shows the observed profiles, the PSF-convolved models, and the corresponding residuals, while the best-fitting structural parameters are summarised in Table~\ref{tab:all_models}.

The pure exponential surface-brightness profile fits provide a stable description of the galaxy in all three bands, with scale lengths $h \sim 32$--$36''$ and extinction-corrected central surface brightnesses $\mu_0 \simeq 19.82$, $19.81$, and $22.35~{\rm mag~arcsec^{-2}}$ in $J$, $H$, and $K_{\rm s}$, respectively. The corresponding face-on values are $\mu_{0,\rm face-on} \simeq 20.91$, $20.91$, and $23.44~{\rm mag~arcsec^{-2}}$. In the independent \texttt{SExtractor}/PSFEx model fits, the low disk-to-total ratio is confined to the $H$ band ($f_{\rm disk}=0.182$), whereas the $J$ and $K_{\rm s}$ solutions are both strongly disk-dominated. This behaviour should therefore not be interpreted as a wavelength-dependent trend, but rather as an instability of the automated $H$-band fit, likely driven by its lower effective depth and/or background structure.

Both approaches support a disk-dominated interpretation, but their scale lengths are not equivalent measurements. The \texttt{SExtractor}/PSFEx values are first-order estimates from automated model fitting, sensitive to segmentation, detection threshold, masking, background treatment, PSF model, and the effective radial extent of the detected light. The direct profile fits explicitly model the azimuthally averaged exponential over the adopted radial range, and their systematically longer scale lengths reflect the more complete treatment of the low-surface-brightness outskirts. We therefore use the \texttt{SExtractor}/PSFEx results as a morphological consistency check and adopt the profile-fit values of $h \sim 32$--$36''$ as the fiducial disk scale lengths.

\section{Summary and conclusions}

We report the discovery of VVVX\,J080705.96$-$273823.6, a low-surface-brightness (LSB) galaxy identified behind the Milky Way using the new VVVX \textit{Deepstack} images. The source is located at $\alpha = 08^{\rm h}07^{\rm m}05\fs96$ and $\delta = -27^\circ38'23\farcs6$ (J2000) and appears as a diffuse and extended system detected in the near-infrared $J$, $H$, and $K_{\rm s}$ images.

Our analysis followed a two-step approach. First, automated modelling with \texttt{SExtractor} indicated disk-dominated solutions in the $J$ and $K_{\rm s}$ bands ($f_{\rm disk}=0.996$ and $0.968$), with a less stable $H$-band fit ($f_{\rm disk}=0.182$) likely affected by the lower effective depth of the image.

The exponential model gives stable scale lengths across the three bands ($h\sim32$--$36''$) and extinction-corrected central surface brightnesses $\mu_0 \simeq 19.82$, $19.81$, and $22.35~{\rm mag~arcsec^{-2}}$ in $J$, $H$, and $K_{\rm s}$, supporting an extended disk-like system.

The projected proximity of VVVX\,J080705.96$-$273823.6 to the nearby galaxies ESO\,494$-$25 and ESO\,494$-$26, located at an angular separation of $\sim 11'$, raises the possibility of a physical association. Both ESO galaxies have radial velocities consistent with distances of $\sim 11$--$12$ Mpc. If VVVX\,J080705.96$-$273823.6 were located at the same distance, its characteristic exponential scale length of $h \simeq 34''$ would correspond to $\sim 1.8$--$2.0$ kpc, consistent with the range of scale lengths observed in nearby LSB disk galaxies \citep{galaz02}. However, in the absence of a spectroscopic redshift, this association remains tentative. Confirming or ruling out a physical connection with the nearby ESO galaxies will require spectroscopic follow-up.

Taken together, its resolved low-surface-brightness morphology, large exponential scale lengths, independent optical confirmation in DECaPS, and central surface brightnesses consistent with those of nearby LSB disk galaxies support a coherent interpretation in which VVVX\,J080705.96$-$273823.6 is a low-surface-brightness disk galaxy dominated by an extended diffuse component. More broadly, the discovery demonstrates that the new VVVX \textit{Deepstack} products can reveal faint diffuse galaxies even in highly obscured regions of the Zone of Avoidance, opening a practical route toward a more complete census of hidden extragalactic structures behind the Milky Way.

\begin{acknowledgements}
J.L.N-C. is grateful to the Universidad de La Serena for providing the academic environment and institutional support that enabled the development of this research. L.D.B., M.V.A. and C.V. thank the support of the Consejo de Investigaciones Cient\'ificas y T\'ecnicas (CONICET) and Secretar\'ia de Ciencia y T\'ecnica de la Universidad Nacional de C\'ordoba (SeCyT). D.M. and G.G. acknowledge support from CATA through ANID BASAL FB210003; D.M. also acknowledges Fondecyt Project No. 1220724 and ANID BASAL ACE210002. M.S. acknowledges support from ANID’s FONDECYT Regular grant No. 1251401. 

We gratefully acknowledge the use of data from the ESO Public Survey program IDs 179.B-2002 and 198.B2004 obtained with the VISTA telescope, and optical data products from the Cambridge Astronomical Survey Unit (CASU), the University of Edinburgh Wide Field Astronomy Unit (WFAU), who are responsible for the VISTA Science Archive (VSA). VVV and VVVX data are published in the ESO Science Archive in the data collections identified by the following DOIs: https://doi.eso.org/10.18727/archive/67 and https://doi.eso.org/10.18727/archive/68. We acknowledge Mike Read for his contribution to the quality control of the VVVX DeepStack images, and Eckhard Sutorius for his major role in the production of the DeepStack products used in this work.
\end{acknowledgements}

\bibliographystyle{aa}  
\bibliography{refs} 
\clearpage
\onecolumn
\appendix
\section{Additional structural models and PSF properties}

\begin{table}[h]
\caption{S\'ersic and composite exponential+S\'ersic PSF-convolved fits. Surface brightnesses (mag\,arcsec$^{-2}$) are corrected for Galactic extinction; scale parameters in arcsec.}
\label{tab:sersic_models}
\centering
\small
\begin{tabular}{lcccccc}
\hline\hline
\multicolumn{7}{c}{\textbf{S\'ersic + PSF}} \\
\hline
Band & $\mu_e$ & $r_e$ & $n$ & $\mu_{0,\rm infer}$ & $\mu_{0,\rm face-on}$ & AIC \\
\hline
$J$  & 22.615 $\pm$ 0.154 & 107.134 $\pm$ 10.445 & 1.821 $\pm$ 0.127 & 19.016 & 20.109 & 390.7 \\
$H$  & 21.715 $\pm$ 0.121 & 56.834 $\pm$ 3.905 & 1.075 $\pm$ 0.099 & 19.738 & 20.830 & 114.5 \\
$K_{\rm s}$ & 26.883 $\pm$ 0.956 & 300.000 $\pm$ 186.941 & 3.295 $\pm$ 0.798 & 20.086 & 21.179 & 184.5 \\
\hline
\multicolumn{7}{c}{\textbf{Exponential + S\'ersic + PSF}} \\
\hline
Band & $\mu_{0,\rm disk}$ & $h$ & $\mu_{e,\rm S}$ & $r_{e,\rm S}$ & $n_{\rm S}$ & AIC \\
\hline
$J$  & 20.159 $\pm$ 0.034 & 45.212 $\pm$ 1.210 & 21.089 $\pm$ 0.082 & 11.128 $\pm$ 0.541 & 0.300 & 266.6 \\
$H$  & 20.029 $\pm$ 0.108 & 36.910 $\pm$ 2.523 & 22.115 $\pm$ 0.330 & 14.207 $\pm$ 2.887 & 0.300 & 108.8 \\
$K_{\rm s}$ & 23.046 $\pm$ 0.095 & 55.642 $\pm$ 4.865 & 23.294 $\pm$ 0.108 & 12.364 $\pm$ 0.841 & 0.300 & 151.1 \\
\hline
\end{tabular}
\end{table}

\setlength{\bibsep}{0pt plus 0.3ex}

\end{document}